# Electron-beam Introduction of Heteroatomic Pt-Si Structures in Graphene

Ondrej Dyck,[1,*] Cheng Zhang,[2] Philip D. Rack,[1,2] Jason D. Fowlkes,[1,2] Bobby Sumpter,[1,3] Andrew R. Lupini,[1] Sergei V. Kalinin,[1] Stephen Jesse[1]

[1] Center for Nanophase Materials Sciences, Oak Ridge National Laboratory, Oak Ridge, TN

[2] Department of Materials Science and Engineering, University of Tennessee, Knoxville, TN

[3] Computational Sciences and Engineering Division, Oak Ridge National Laboratory, Oak Ridge, TN

**Abstract**

Electron-beam (e-beam) manipulation of single dopant atoms in an aberration-corrected scanning transmission electron microscope is emerging as a method for directed atomic motion and atom-by-atom assembly. Until now, the dopant species have been limited to atoms closely matched to carbon in terms of ionic radius and capable of strong covalent bonding with carbon atoms in the graphene lattice. *In situ* dopant insertion into a graphene lattice has thus far been demonstrated only for Si, which is ubiquitously present as a contaminant in this material. Here, we achieve *in situ* manipulation of Pt atoms and their insertion into the graphene host matrix using the e-beam deposited Pt on graphene as a host system. We further demonstrate a mechanism for stabilization of the Pt atom, enabled through the formation of Si-stabilized Pt heteroatomic clusters attached to the graphene surface. This study provides evidence toward the universality of the e-beam assembly approach, opening a pathway for exploring cluster chemistry through direct assembly.

## 1. Introduction

Atom-by-atom assembly of materials and devices is a long-standing target of nanometer-scale science and technology. Classical synthesis science has achieved remarkable progress in the synthesis of complex molecular and cluster species *en masse* in addition to well-developed pathways for computer-controlled and intelligent synthesis.[1-2] Meanwhile, the broader imaging communities, including electron microscopy and scanning probe microscopies (SPM), have demonstrated the ability to conduct high-resolution imaging of atomic and molecular structures routinely. By connecting these developments, a number of SPM groups have demonstrated methods for using the SPM tip to desorb atomic species, induce lateral atomic motion, and to introduce complex chemical reactions.[3-12] This approach was used to explore the fundamental mechanisms of complex reaction processes and is now being used as a basis for single-atom device fabrication, most notably P/Si qubits.[13-16]

Recently, similar opportunities have been realized using the scanning transmission electron microscope (STEM). Observations of electron-beam (e-beam)-induced atomic-level changes in a number of materials have prompted the application of e-beam manipulation as a way to engineer atomic-scale structures in solids.[17] Researchers are currently exploring e-beam manipulation of single dopant atoms in graphene; the controlled *in situ* introduction[18-20] and positioning[19, 21-23] of Si dopants via directed e-beam manipulation has been demonstrated as well as *ex situ* methods for introducing other dopant species including P,[24-25] B, N,[26] and Ge.[27] In these latter experiments, dopants were introduced into the graphene either during growth or implanted through ion irradiation. Notably, these elemental species are chemically similar to carbon, e.g., have close ionic radii and are capable of forming strong covalent bonds with the carbon atoms. The spectrum of dopants in graphene that have been successfully manipulated *in situ* is much smaller and to date the only dopant species that has been intentionally introduced into the lattice *in situ* using the e-beam has been Si.[18-19] Interestingly, the atomic manipulation of Si on graphene was recently extended toward complex cluster assembly, where Si dimers, trimers, and tetramers were assembled with an e-beam.[19] These developments naturally lead to questions as to what atomic species can be used in e-beam manipulation and assembly and the degree to which such atomic-level assembly can be integrated with semiconductor workflows.

Here, we explore the manipulation of Pt atoms deposited onto graphene surface contamination. We demonstrate two methods for inserting the Pt atoms into the graphene lattice: positioning a Pt atom at the edge of a milled hole formed in the graphene lattice and the two-step formation involving 1) the insertion of Si dopant(s) and 2) the attachment of a Pt atom to form a heteroatomic dopant cluster. The assembly of heteroatomic structures accomplishes the next level of complexity in atom-by-atom fabrication for 2D materials.[28]

## 2. Results

As a first step, we explored various pathways for the introduction of Pt dopant atoms onto the graphene lattice. We began by depositing Pt via focused e-beam-induced deposition (FEBID) of trimethyl(methylcyclopentadienyl) platinum(IV) (MeCpPtMe$_3$) precursor gas. The deposition was chosen so that the creation of Pt contact lines for future device applications and deposition of Pt dopant atoms for manipulation could be investigated simultaneously. Pt lines were directly deposited onto the graphene sample and were imaged in the STEM to observe the Pt morphology. Figure 1a shows a high angle annular dark field (HAADF)-STEM image of one such set of Pt deposition lines (see methods section). Figure 1b shows an area ~1 μm away from the deposited lines where Pt nanoparticles and dispersed single Pt atoms were observed in the ubiquitous contamination typically observed on graphene surfaces. Figure 1c shows the Pt $O_{23}$ edge for the electron energy loss spectrum (EELS) acquired on a Pt nanoparticle (inset).

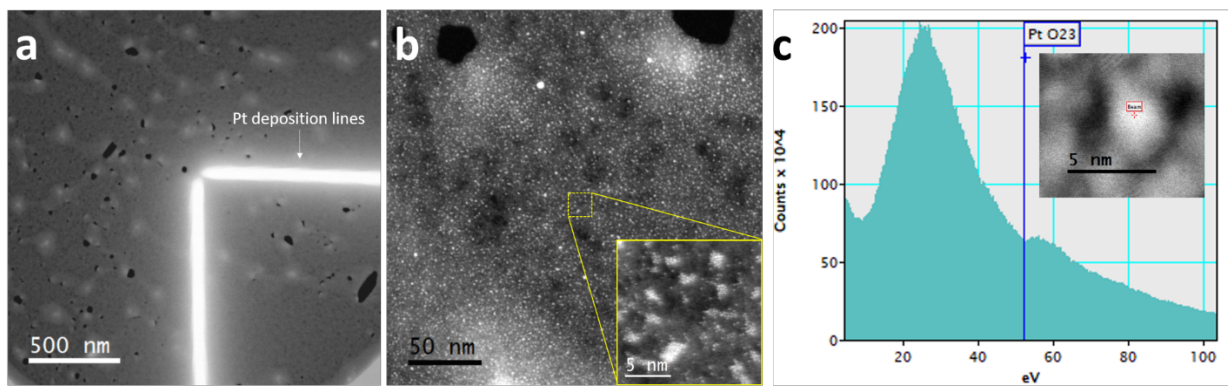

**Figure 1 Overview of Pt deposition on graphene and surrounding sample.** a) HAADF-STEM image of Pt deposition lines on top of suspended graphene. b) Area ~1 μm away from Pt deposition lines shows Pt nanoparticles and single Pt atoms were also deposited during relatively long line exposure times used to deposit Pt lines. Inset shows higher magnification image of the central region. c) EELS Pt $O_{23}$ edge observed with e-beam positioned on Pt nanoparticle shown (inset).

Interestingly, Pt nanoparticles and single atoms were found on all contamination areas of the sample investigated, even areas far from the Pt line deposition sites. This deposition of outlying Pt is attributed to scattered electrons that induce dissociation of the MeCpPtMe$_3$ precursor far from the intended deposition sites. Since the areas near the Pt line deposition sites were more heavily covered with Pt and other contamination from the precursor gas, regions far from the intended deposition site were used for our experiments where both Pt atoms (on contamination areas) as well as clean graphene co-exist.

As a first pathway towards Pt atomic manipulation, we explored the "sputter and damage" method that has been previously demonstrated for the manipulation of Si dopants.[19] Here, the adjacent pristine graphene lattice is scanned with the e-beam concurrently with the source material containing the element of choice, which results in the introduction of dopants.

This method relies on the creation of e-beam-induced vacancies/defects in the pristine graphene lattice while potential dopant atoms are sputtered from the source material and diffuse across the graphene surface. Figure 2 shows the result of this method when the source material (in this case, an adjacent layer of contamination) only contains a few Pt atoms. Figure 2a shows a HAADF-STEM image of the initial configuration where the bright areas are the surface contamination layers (source material containing Pt) and the dark regions are pristine graphene. The yellow arrow highlights that initially (frame 1) there are no Pt atoms observable within the pristine area of graphene. After scanning a few frames over the source material and the graphene a bright Pt atom has moved onto the clean graphene region, as shown in Figure 2b. A second similar example is demonstrated in Figure 2c and 2d. Figure 2e shows a HAADF-STEM image of a single Pt atom. From this image it is not clear whether the Pt atom has been inserted into the graphene lattice as a substitutional dopant or is bonded to the surface, possibly associated with a defect location. Based on lattice symmetry, the location of the Pt atom is consistent with the occupation of a graphene divacancy, illustrated in the overlay. The Pt atom is, however, slightly offset from the expected position and the lattice below and to the right of the Pt atom does not align well to the hexagonal overlay, suggesting that some lattice distortion is responsible. Additionally, the lattice in this region appears slightly brighter than the rest of the graphene lattice which may be due to additional atoms loosely bound to the surface which are interacting with the Pt atom (e.g. a hydrocarbon chain).

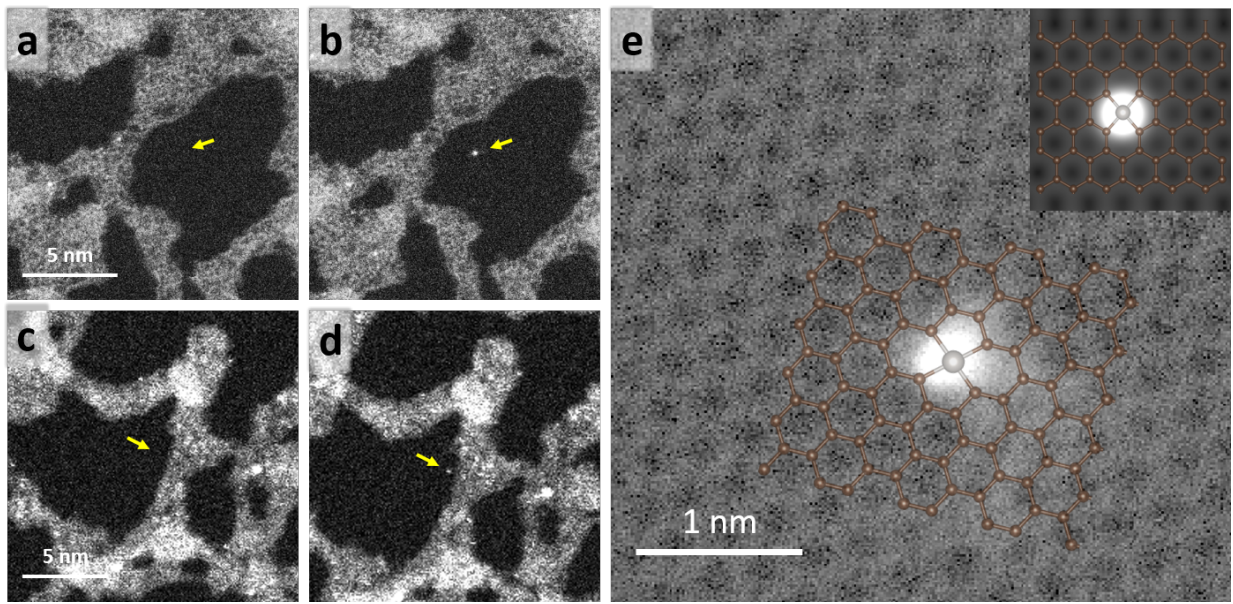

**Figure 2 Moving Pt atoms on graphene by scanning the e-beam.** a) HAADF-STEM image of initial configuration of patches of pristine graphene within a network of amorphous C/Si contamination layers containing a few Pt atoms. b) After scanning 100 kV e-beam several times, a Pt atom becomes either attached to the graphene surface (likely at a defect site) or occupies a lattice vacancy. c) and d) show a second example of a similar process demonstrated in a) and b). e) HAADF-STEM image of Pt atom acquired at 60 kV with a structure model overlay. Inset is a multislice simulated image the suggested structure showing qualitatively similar appearance to the experimental image.

As a result of the high scattering cross-section of the Pt atom relative to the graphene lattice, the graphene structure immediately adjacent to the Pt atom cannot be resolved in the HAADF-STEM images in Figure 2. Thus, to explore possible bonding configurations between the Pt atom and the carbon atoms in the graphene lattice, density functional theory (DFT) simulations were performed to compare several potential structures including a Pt atom attached to a single carbon vacancy (Pt@SV) or a double carbon vacancy (Pt@DV). Figure 3 summarizes the DFT analyses of these structures as well as others informed by experimental observations (discussed later). The Pt@SV structure, shown in plan-view in Figure 3a and in cross-section in Figure 3b, has a binding energy of 7.7 eV and the configuration protrudes from the graphene basal plane by 1.68 Å. In this configuration the surrounding graphene lattice is distorted, as illustrated in Figure 3c. A slightly more stable structure is afforded by Pt@DV, as shown in Figure 3d and 3e. This structure has a binding energy of 7.8 eV and the geometrically larger double vacancy space allows the Pt atom to reside within the basal plane with a longer Pt-C bond length of 1.95 Å and no appreciable distortion in the surrounding lattice. Figure 3f and 3g show a stable structure that involves both a Pt and Si dopant occupying a tri-vacancy cluster (Pt-Si@TV) in the graphene. For the Pt-Si@TV configuration the Pt binding energy is much lower, 3.9 eV; however, this structure showed a stronger binding energy of 5.3 eV when considering both the Pt and Si together. Figure 3h and 3i show the plan-view and cross-sectional view, respectively, of a structure formed by the removal of four carbon atoms, the subsequent addition of a silicon trimer into the vacancy, and the addition of a Pt atom on top of the Si trimer. Figure 3j shows a plot of the Z-height distortion of the graphene as a function of distance from the Pt atom. The Pt atom is less stable in this configuration than when it occupies a single carbon vacancy or divacancy and has a binding energy of 5.7 eV. Figure 3k and 3l show the plan-view and cross-sectional view, respectively, of a Pt-Si defect cluster created by removing six carbon atoms, attaching four Si atoms to the inside of the created pore, and attaching a Pt atom to the center of the four Si atom defect cluster. This configuration produces significant distortion in the surrounding graphene lattice as shown in the Z-height distortion plot of Figure 3m; nevertheless, the Pt binding energy of 8.5 eV is the highest of all the structures investigated here, indicating that more complex structures and larger pores lined with Si atoms introduce a significant stabilizing effect.

In contrast, a Pt adatom on a graphene surface is bound to carbon by only 1.97 eV and the diffusion barrier for a Pt atom to migrate along the surface is ~0.14 eV,[29] which is sufficiently low that diffusion is expected at room temperature. Pt attached to a topological defect (5-7-5 member rings) shows a slightly higher binding energy of 2.55 eV (not shown).[30] During the acquisition of a HAADF-STEM image (as shown in Figure 2e, for example), on the order of four million electron impacts are delivered to the Pt atom without inducing any instability and thus, it is likely there is an underlying defect in the graphene acting as a stabilizer. Moreover, the defect sites created in the graphene lattice by the 100 kV e-beam (and the fact that the e-beam imparts energy) will cause the Pt adatoms scattered across the graphene surface to have a strong propensity to attach to such defect sites. Once attached, the migration barrier jumps to 3.1 eV for Pt@SV and ~5 eV for Pt@DV.[31] Similar increases in the migration barrier for other the other types of defects shown in Figure 3 is also expected; therefore, the most likely observable configurations under the conditions presented herein are Pt atoms bound to a range of potential defect sites.

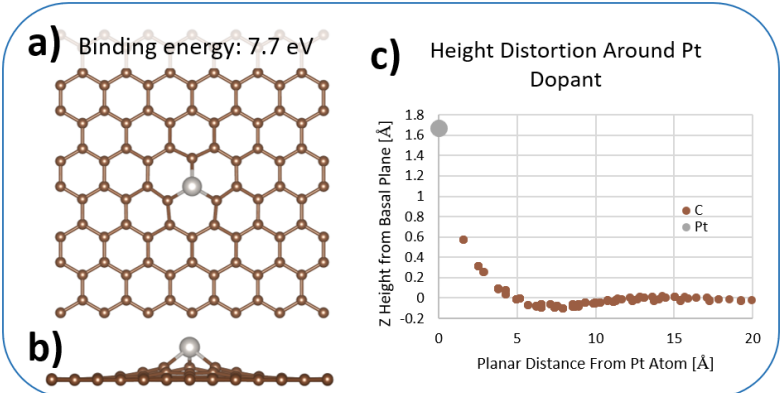
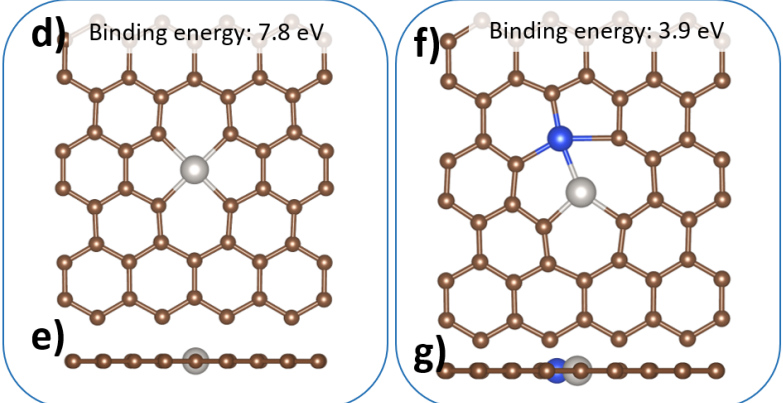
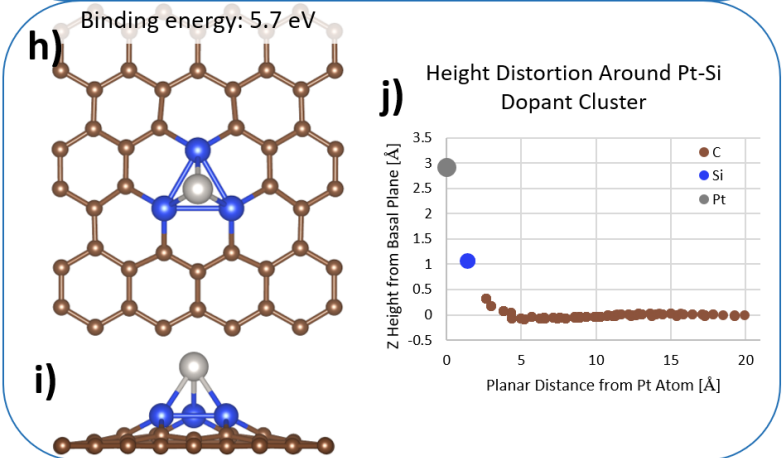
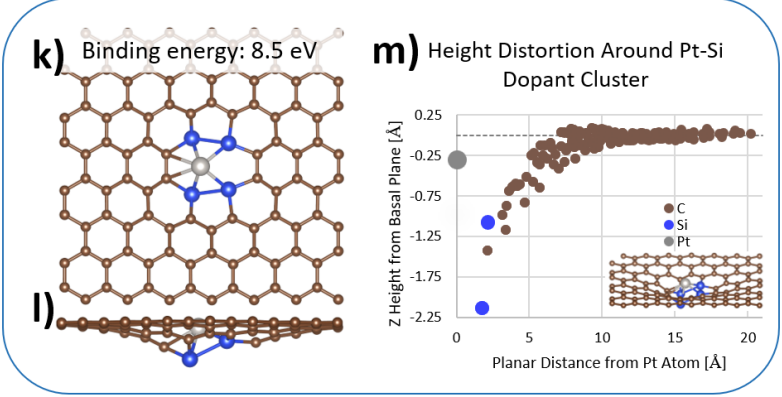

**Figure 3 DFT simulated structures of stable Pt and Pt:Si defect configurations in graphene.** Structures formed for Pt atom occupying a graphene vacancy (Pt@SV): a) plan-view rendering of converged structure - Pt atom has a binding energy of 7.7 eV; b) corresponding cross-sectional view of structure in a); c) plot of the Z-height of each atom above graphene basal plane as a function of planar distance from Pt atom position highlighting degree of vertical distortion around defect. d)-e) Plan-view and cross-sectional view of structure of Pt atom occupying a graphene divacancy (Pt@DV).- Pt binding energy is 7.8 eV. f)-g) Plan-view and cross-sectional view of Pt-Si dimer occupying a graphene tri-vacancy (Pt-Si@TV) - Pt binding energy is 3.9 eV; however Pt-Si@TV showed enhanced binding energy of 5.3 eV over Pt alone. h)-j) Structure where a Si trimer replaces four carbon atoms and a Pt atom binds to the top of this trimer: h) plan-view and i) cross-sectional view - binding energy for Pt atom in this configuration is 5.7 eV; j) plot of Z-height of each atom above graphene basal plan as a function of planar distance from Pt atom position highlighting degree of vertical distortion around defect. k)-m) More complex defect structure involving four Si atoms in the Pt-Si defect cluster, which occupies a pore resulting from the removal of six carbon atoms from graphene lattice: k)-l) plan-view and cross-sectional view of optimized structure - Pt binding energy is 8.5 eV; m) plot of Z-height of each atom above graphene basal plane as a function of planar distance from Pt atom position highlighting degree of vertical distortion around defect cluster. Inset is rendering of structure at an angle to better visualize distortion.

A more targeted approach to introduce a Pt atom into a graphene defect site is shown in Figure 4. Here, a Pt atom observed at the edge of the contamination layer and pristine graphene was chosen for e-beam manipulation (Figure 4a). The e-beam was initially positioned next to the Pt atom on the graphene as indicated by the yellow spot shown in Figure 4b. A small hole was created due to e-beam-induced sputtering of the carbon atoms. (Figure 4c). To prevent further e-beam damage to the graphene lattice and to promote healing of the hole, the e-beam was manually moved to scan across a region of the adjacent contamination layer, as indicated by the yellow box shown in Figure 4d. Since this region contained amorphous carbon contamination on the graphene surface, sputtering during e-beam irradiation served to eject additional carbon atoms from the contamination layer, some of which were transferred toward the pore and contributed to healing, as demonstrated in Figure 4e. On close inspection of the image, it appears the Pt atom remained bound to a few Si atoms from the contaminated area and the whole cluster of Pt-Si atoms migrated onto and attached to the graphene lattice. A few similar examples of this type of structure are shown Figure 4k-m. The suggested structures are displayed on the right with associated multislice simulated images to compare to the experimental data and determine qualitatively how well the suggested structure reproduced the image.

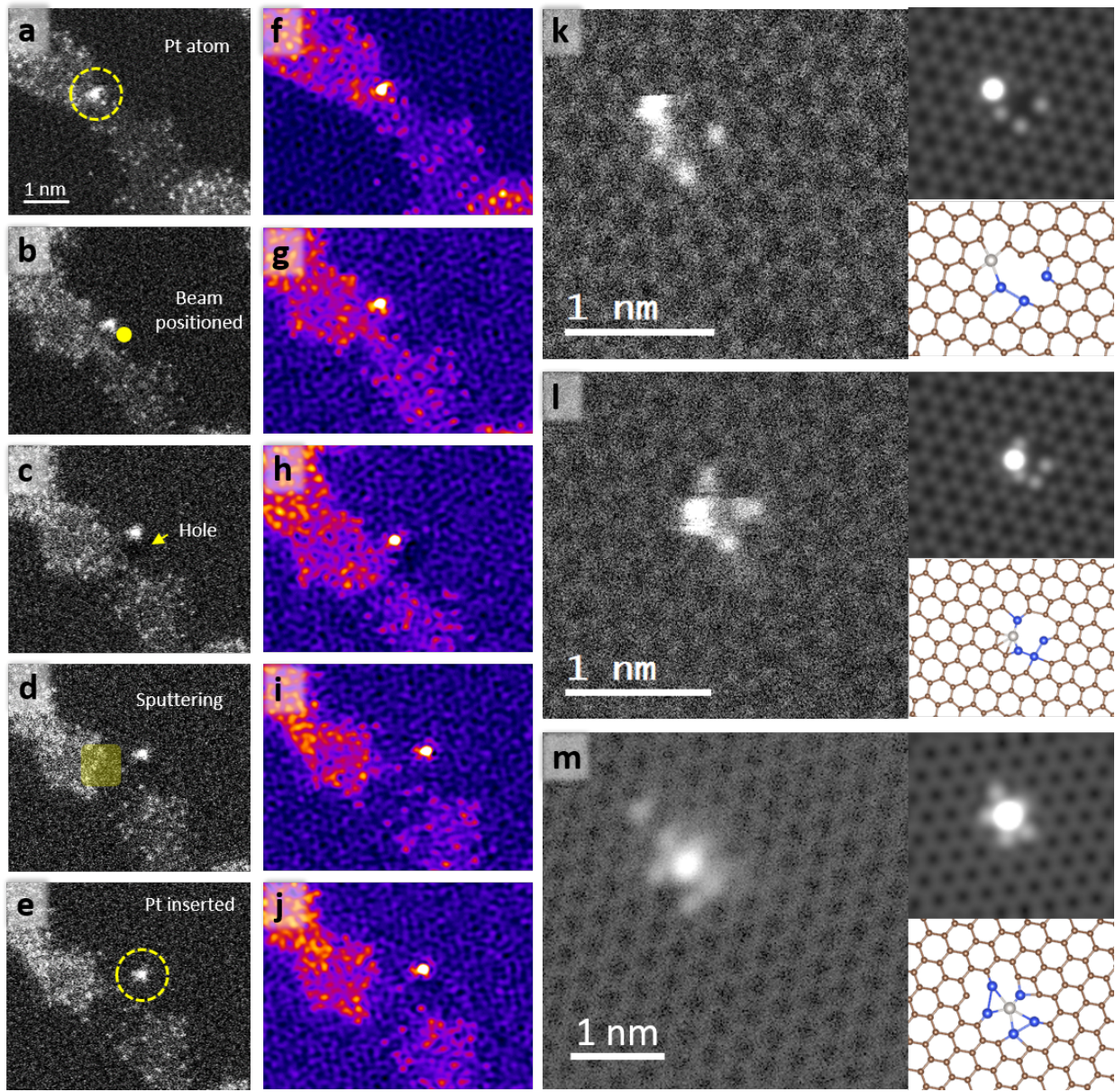

**Figure 4 Insertion of a Pt atom into defect in graphene lattice.** a) Initial configuration with a single Pt atom (brightest) at edge of contaminated region next to area of pristine graphene. b) e- beam positioned at location in graphene adjacent to Pt atom (yellow dot). c) Small pore created through e-beam-induced sputtering of C atoms from graphene lattice that allowed Pt atom to attach to edge of pore. d) e-beam manually moved over yellow shaded area to sputter more carbon atoms to pore allowing it to heal and move contaminated region away from the Pt atom. e) Final configuration of Pt atom in graphene lattice. Close inspection reveals that several brighter atoms (possibly Si) remain adjacent/attached to the Pt atom. f)-j) Cleaned versions of images in a)-e). Images cleaned using Pycroscopy,[32] artificially colored with "fire" look up table (LUT) in Fiji.[33] k)-m) show HAADF-STEM images of additional examples at higher magnification of Pt-Si clusters with multislice simulated images and suggested structure models on the right. HAADF-STEM image in m) acquired at 60 kV; others at 100 kV.

Here, it should be noted that the defect clusters being examined are continuously being influenced by the e-beam during the imaging process. This leads to dynamic structural reconfigurations of the defect clusters as evidenced by streaking and "broken" intensities from line to line in Figure 4k and l. Specifically, the intensity onset of the Pt atom in k is unphysically sharp, suggesting that the Pt atom moved to the observed location during the image acquisition (i.e. at the moment when the intensity became bright). Similarly, the observed region of intensity does not appear round in this image, suggesting movement between two stable minima during the acquisition. In l, streaking is also observed on the Pt and Si atoms suggesting significant movement during the imaging process. The suggested atomic models fail to capture these dynamic processes and, particularly in l, we can see that the simulated image does not match the observed intensity. The Pt and Si atoms have changed position dynamically within the graphene pore which the atomic model cannot capture. The image in m was acquired at a lower accelerating voltage (60 kV) which imparts less energy to the sample during imaging and likely accounts for why we do not observe similar streaking in this image. To illustrate these described dynamics more clearly, a 50-frame video is provided in the supplemental information capturing an example of the Pt-Si dynamic restructuring under the influence of the 100 kV e-beam. A 2D gaussian blur was used to reduce noise and the video was artificially colored using the "fire" look up table (LUT) in Fiji.[33]

During e-beam manipulation, we observed that the Pt atoms had a strong propensity toward interacting (bonding) with other localized atomic species present in the contamination layer, as shown in Figure 5. In this example, a Pt atom was bonded to an unknown lighter atom, which stretched the circular shape of the atom into an oblong shape. This cluster resided on a section of bilayer graphene and the e-beam was positioned to induce sputtering of the graphene step edge shown in Figure 5a. It should be noted that the two brighter atoms toward the bottom of the image are neither Pt nor Si, based on their observed intensities. Figure 5b shows the result of this step edge sputtering, where the atom cluster became attached to both the step edge and a pore that was created in the adjacent pristine graphene. The e-beam was then positioned on the other side of the atom cluster to sputter away the bilayer graphene in this area, which resulted in the configuration shown in Figure 5c where the pore at the step edge has been enlarged and the atom cluster became attached to the edge of the pore in the pristine graphene, Figure 5d. DFT simulations of a Pt atom attached to the edge of a pore in graphene revealed a strong binding energy of 9.79 eV with a low migration barrier for movement along the edge. Molecular dynamics (MD) simulations were carried out to reveal this mobility; a video of the simulated behavior can be found in the supplemental materials.

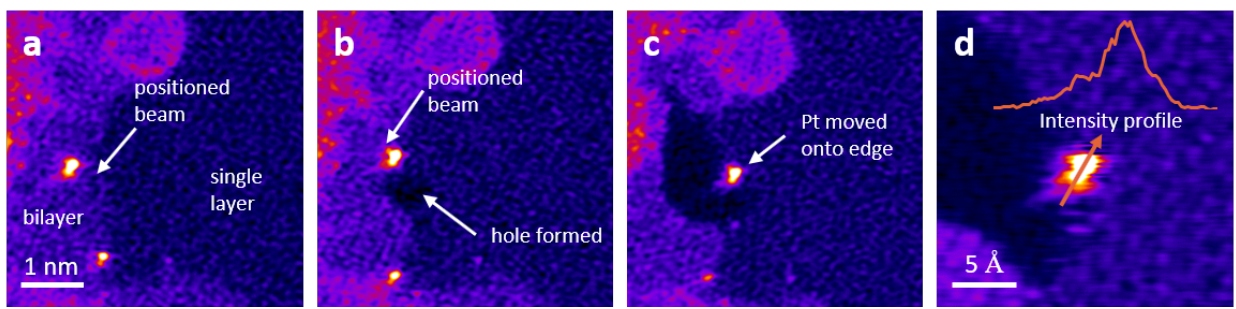

**Figure 5 Attaching a Pt atom to the edge of a pore in graphene.** a) HAADF-STEM image of initial configuration of the sample. Pt atom resides near step edge between single and bilayer graphene. Oblong shape indicates there is another bright atom bonded to the Pt. It is unclear what element this is. e-beam was positioned as indicated by arrow and a small pore formed as shown in b). e-beam then used to sputter atoms adjacent to Pt atom to attach to edge of pore, which increased pore size as shown in c); Pt atom successfully attached to pore edge. d) Higher magnification HAADF-STEM image of final configuration. An intensity profile from raw data shown inset to show difference in brightness of two atoms. All images cleaned using Pycroscopy[32] and artificially colored using "fire" LUT in Fiji.[33]

To explore this mobile atom behavior further, we attempted to use the e-beam to direct a Pt atom across the graphene surface in the presence of other Si dopant atoms, as demonstrated in Figure

6. Here, an e-beam dragging technique was used where the focused e-beam was manually positioned and moved across the sample between acquisition of the HAADF-STEM images. By repeatedly dragging the e-beam from the source material (e.g., the contamination layer) comprised mostly of amorphous C interspersed with Si atoms and a few Pt atoms (see supplemental information) across to the pristine graphene, the atoms of the source material could be sputtered over the adjacent graphene surface. The e-beam simultaneously created vacancies and other defects in the graphene lattice that could serve as attachment sites for the sputtered atoms. Figure 6a shows the initial stage of this process where a few Si atoms have been sputtered (moved) into the graphene lattice. As the process continued, the e-beam was used to specifically target the Pt atoms, which resulted in the sequential configurations shown in Figure 6b and Figure 6c; a Pt atom sputtered from the source material attaches to several of the inserted Si dopants. A higher exposure image (longer dwell time) was attempted, as shown in Figure 6d, but the Pt atom was knocked away by the e-beam.

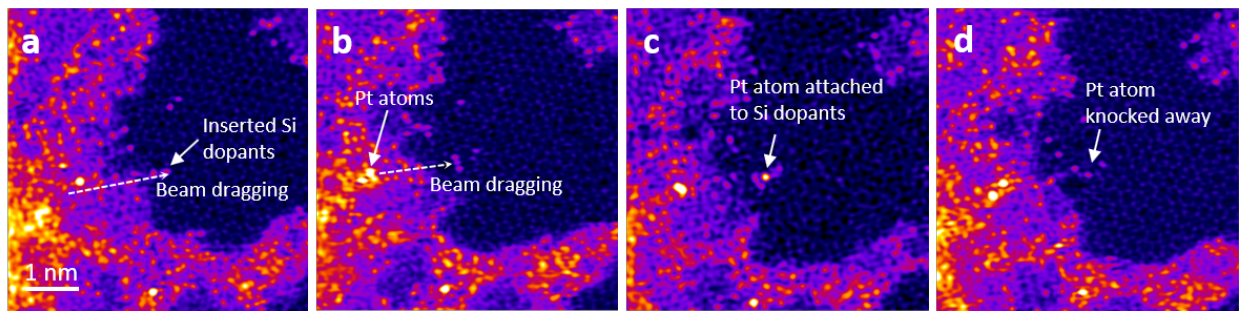

**Figure 6 Inserting Si atoms into the graphene lattice and attaching a Pt to the dopant cluster.** a) By manually positioning and dragging e-beam from source material over the pristine graphene, Si dopants were introduced into the generated lattice defect sites. b) Continued e-beam dragging over Pt atoms toward inserted Si dopants results in c), a Pt atom attaching to dopant cluster. Upon imaging with a slower scan d), Pt atom was knocked away indicating a low energy barrier for removal. All images were cleaned using Pycroscopy[32] image cleaning tools and artificially colored using the "Fire" look up table (LUT) in Fiji.[33]

We observed the potential attachment of Pt atoms to defect sites formed in the graphene lattice (Figure 2), which was validated by DFT simulations; however, most of the examples involve the formation of a heteroatomic defect cluster where Si atoms form a buffer/interlayer between the graphene lattice and the Pt atom(s). This is likely due to a greater affinity for Pt to bond with Si compared with C[34] as well as a propensity for Si dopants to be inserted into the graphene lattice during e-beam irradiation,[19] thereby acting as anchoring sites for Pt atoms and stabilizing the defect configuration.[35-36] DFT calculations of a co-inserted Si with Pt into a tri-vacancy, Pt-Si@TV, showed a binding energy of 5.3 eV (Figure 3f and 3g). A stronger binding energy of 8.5 eV for the Pt atom was found by creating a larger hole in the graphene lattice ringed with Si atoms. The strategy of using a buffer element for inserting dopants into the graphene lattice may enable unstable elements to potentially be held in position at designated graphene lattice defects (Figure 3h-m).

3. Discussion

Recent work demonstrates that control of the e-beam in STEM can be used to manipulate single dopant atoms in both 2D and 3D materials; however, to ultimately create structures for practical applications, several additional steps are necessary. For the controlled e-beam-induced incorporation of dopants into a lattice using the techniques demonstrated by Dyck[18-19] and manipulation methods advanced by Susi[22-23, 37] to be useful, the insertion of a range of different (specified) atoms and precise control over where they are positioned will be required; in addition, there must be a viable means to establish electrical contact. Platinum is of widespread interest for catalytic applications and thus, validating that Pt can be strategically incorporated into a crystalline lattice via such methods provides a viable route toward chemical functionality for devices constructed atom-by-atom. The instability and motion of the Pt atoms during our experiments, which were conducted at 100 and 60 kV, suggest that working at different accelerating voltages, variable dose conditions, and/or incorporating precise scan control methods will be required.

For electronic devices or for proof-of principle electrical measurements, electrical contacts must be close to the area of interest, which becomes extremely challenging for atomic-scale devices. Here we have demonstrated that although the peripheral contamination associated with FEBID deposition can be a potentially limiting factor for introducing elements onto graphene, the low density of metal atoms dispersed across on the sample provide a good source of single dopants with some control over their locations. We anticipate that the combined use of FEBID and atomic-resolution STEM could provide a source of single atoms of other metals as well as providing a method to connect single atom structures to the outside world.

## 4. Conclusions

In this study, we explored pathways for introducing and manipulating Pt atoms in the graphene lattice. Unlike previously explored dopants such as Ge, Si, B, P, and N, Pt has considerably dissimilar chemical properties and bonding with carbon and other dopant elements such as Si. Furthermore, Pt is considerably heavier than previously attempted dopants.

Here, we demonstrated the *in situ* manipulation of Pt atoms with a focused e-beam in a STEM. Pt atoms were deposited onto contamination layers on the surface of graphene via e-beam induced deposition. Manipulation through controlled e-beam exposure was demonstrated using either a targeted approach where a specific Pt atom was identified and selectively nudged as well as a more random approach where the e-beam was scanned over a selected area (that included a contamination layer and pristine graphene) to encourage the Pt atoms to scatter across the graphene lattice and attach to defect sites. A range of possible configurations were explored using DFT methods. Creation of a pore and the targeted attachment of a Pt atom (bound to another unidentified atom) to the edge of the pore was also shown. Finally, using an e-beam dragging technique, we showed the two-step process of inserting Si dopants into the graphene lattice followed by the attachment of a Pt atom to these dopants.

These studies suggest that e-beam atomic manipulation can potentially be applied to a broader spectrum of atomic species than previously assumed based on chemical similarity and knock-on thresholds. Finally, we note that these approaches allow for atom-by-atom fabrication of complex atomic clusters with graphene (and potentially other substrates) as a stabilizing matrix. Subsequent probing of their functional properties may also be possible.

## 5. Methods

### 5.1. Sample Preparation

Graphene was grown on Cu foil via atmospheric pressure chemical vapor deposition (AP-CVD) followed by spin coating of poly(methyl-methacrylate) (PMMA) to form a mechanically stabilizing layer for wet transfer and to serve as a barrier to corrosion during storage. An approximately 9 mm$^2$ square was cut from the main foil and set afloat on a petri dish containing an ammonium persulfate-deionized (DI) water solution (0.05 g/ml of ammonium persulfate). The Cu foil was allowed to dissolve and the remaining graphene/PMMA stack was removed and set to float on clean DI water to allow the ammonium persulfate solution to wash away. The graphene/PMMA stack was then scooped up using a TEM grid and baked on a hot plate at 150 °C for ~20 minutes to promote adherence of the graphene to the grid. The whole grid was then submerged in a bath of acetone and gently swished back and forth for a minute or two to dissolve the PMMA. The grid was then rinsed with isopropyl alcohol (IPA) before the acetone dried. While still wet with IPA the grid was set on filter paper and allowed to dry. The grid was then transferred to a tube furnace and baked at 500 °C for 1.5 hrs under an Ar/O$_2$ (450 sccm/50 sccm) environment to mitigate hydrocarbon deposition in the microscope.[38] Prior to loading the sample in the microscope the sample cartridge (with sample) and magazine were baked under vacuum at 160 °C for 8 hours.

Pt rods/lines were deposited on the graphene sample via focused electron-beam-induced deposition (FEBID) using a FEI Nova 600 dual beam electron/ion beam microscope. The MeCpPt$^{IV}$Me$_3$ precursor crucible used as the Pt source was heated to 45 °C for 5 min. and allowed to flow through the gas injection nozzle placed in close proximity (~ 100 μm in x and z from the electron impact region) before deposition. During deposition, the electron energy was set at 5 kV with a beam current of 40 nA, and the dwell time to deposit the Pt lines varied from 0.5 to 5 ms to control the rod/line thickness.[39]

### 5.2. Imaging

Scanning transmission electron microscopy (STEM) characterization and e-beam manipulation was performed in a Nion UltraSTEM US200 at an accelerating voltage of 100 kV (60 kV for the image in Figure 4m) with a beam current of ~17 pA (~20 pA). Electron energy-loss spectroscopy (EELS) was performed at 100 kV accelerating voltage with a convergence semi-angle of 30 mrad and a collection semi-angle of 33 mrad with a total integration time of 5 s. A dark reference was collected immediately after acquisition and subtracted from the spectrum to correct for camera nonuniformities.

### 5.3. Image Simulation

Multislice HAADF image simulations were performed using the Dr. Probe software package.[40] The frozen phonon method was used with 64 displacements per atom and 10 variations per pixel. The sampling rate was approximately 0.02 nm/pixel. The convergence angle was set to 30 mrad and the detector collection angle spanned 80-250 mrad. Source sizes were chosen such that the simulated image appeared qualitatively similar to the experimental image. These ranged from 0.045 to 0.07 nm depending on the image.

### 5.4. Density Functional Theory (DFT)

The spin polarized DFT calculations were carried out using the Vienna *Ab Initio* Simulation Package (VASP, 5.4.1)[41-42] using the projector-augmented wave (PAW) method[43-44]. The electron-ion interactions were described using standard PAW potentials, with valence electron configurations. A kinetic energy cutoff on the plane waves was set to 400 eV and the "accurate" precision setting was adopted. A Γ-centered k-point mesh of 9x9x1 was used for the Brillouin-zone integrations. The convergence criteria for the electronic self-consistent loop was set to $10^{-4}$ eV. As mentioned, the binding energies for different Pt – graphene geometries were calculated as the difference in energy between a Pt atom in the substitutional position and a vacancy/vacancies and an isolated Pt atom; similarly, for cases for Si co-substitution. Room temperature ab initio dynamics was also carried out to check the thermal stability of the Pt-graphene materials. Following full optimization MD simulations were performed in a canonical NVT ensemble with the temperature of the simulations maintained by a Nose-Hoover thermostat at 25 C. A time step of 1 fs was used with the velocities of all atoms initialized randomly according to the Maxwell-Boltzmann distribution. The dynamics was evaluated over 100 ps.

**Acknowledgements**

This work was supported by the U.S. Department of Energy, Office of Science, Basic Energy Sciences, Materials Science and Engineering Division (O.D., A.L., S.V.K, B.S., S.J.) and was performed at the Oak Ridge National Laboratory's Center for Nanophase Materials Sciences (CNMS), a U.S. Department of Energy, Office of Science User Facility. P.D.R. and J.D.F. acknowledge support for the electron beam induced deposition was provided by the Nanofabrication Research Laboratory at the Center for Nanophase Materials Sciences, which is a DOE Office of Science User Facility. CZ acknowledges support from the US Department of Energy (DOE) under Grant No. DOE DE-SC0002136.